\input harvmac.tex
\vskip 1.5in
\Title{\vbox{\baselineskip12pt 
\hbox to \hsize{\hfill}
\hbox to \hsize{\hfill CTP-SCU/2015021 }}}
{\vbox{
	\centerline{\hbox{Higher Spins at the Quintic Order:
		}}\vskip 5pt
        \centerline{\hbox{Localization Effect and Simplifications 
		}} } }
\centerline{Dimitri Polyakov$^{}$\footnote{$^{\dagger(1),(2)}$}
{polyakov@sogang.ac.kr ; polyakov@scu.edu.cn; 
twistorstring@gmail.com
}}
\medskip
\centerline{\it Center for Theoretical Physics $^{(1)}$}
\centerline{\it College of Physical Science and Technology}
\centerline{\it Sichuan University, Chengdu 610064, China}
\centerline{\it }
\centerline{\it }
\centerline{\it Institute for Information Transmission Problems (IITP)$^{(2)}$}
\centerline{\it Bolshoi Karetny per. 19/1}
\centerline{\it 127994 Moscow, Russia}
\vskip .3in

\centerline {\bf Abstract}

We investigate the special case of quintic interactions for massless higher spin gauge fields
using the string-theoretic vertex operator construction for higher spin gauge fields in Vasiliev's frame-like
formalism. We compute explicitly the related 5-point interaction vertex
in the low energy limit of string theory and find that:
 the structure of the quintic $s_1-s_2-s_3-s_4-s_5$ higher 
spin interaction gets drastically simplified
and localized if

a) the spin values satisfy the constraint $s_1+s_2+s_3=s_4+s_5+2$ (and, more generally, if the sum
of three spin values roughly equals the sum of the remaining two

b) One of the spin values, $s_4$ or $s_5$ is sufficiently small.
In this paper, the explicit computation is done for the case $s_4=4$

\Date{November 2015}

\vfill\eject
\lref\sagnottia{A. Sagnotti, E. Sezgin, P. Sundell, hep-th/0501156}
\lref\sorokin{D. Sorokin, AIP Conf. Proc. 767, 172 (2005)}
\lref\fronsdal{C. Fronsdal, Phys. Rev. D18 (1978) 3624}
\lref\coleman{ S. Coleman, J. Mandula, Phys. Rev. 159 (1967) 1251}
\lref\haag{R. Haag, J. Lopuszanski, M. Sohnius, Nucl. Phys B88 (1975)
257}
\lref\weinberg{ S. Weinberg, Phys. Rev. 133(1964) B1049}
\lref\fradkin{E. Fradkin, M. Vasiliev, Phys. Lett. B189 (1987) 89}
\lref\skvortsov{E. Skvortsov, M. Vasiliev, Nucl.Phys.B756:117-147 (2006)}
\lref\skvortsovb{E. Skvortsov, J.Phys.A42:385401 (2009)}
\lref\mva{M. Vasiliev, Phys. Lett. B243 (1990) 378}
\lref\mvb{M. Vasiliev, Int. J. Mod. Phys. D5
(1996) 763}
\lref\mvc{M. Vasiliev, Phys. Lett. B567 (2003) 139}
\lref\brink{A. Bengtsson, I. Bengtsson, L. Brink, Nucl. Phys. B227
 (1983) 31}
\lref\deser{S. Deser, Z. Yang, Class. Quant. Grav 7 (1990) 1491}
\lref\bengt{ A. Bengtsson, I. Bengtsson, N. Linden,
Class. Quant. Grav. 4 (1987) 1333}
\lref\boulanger{ X. Bekaert, N. Boulanger, S. Cnockaert,
J. Math. Phys 46 (2005) 012303}
\lref\metsaev{ R. Metsaev, arXiv:0712.3526}
\lref\siegel{ W. Siegel, B. Zwiebach, Nucl. Phys. B282 (1987) 125}
\lref\siegelb{W. Siegel, Nucl. Phys. B 263 (1986) 93}
\lref\nicolai{ A. Neveu, H. Nicolai, P. West, Nucl. Phys. B264 (1986) 573}
\lref\damour{T. Damour, S. Deser, Ann. Poincare Phys. Theor. 47 (1987) 277}
\lref\sagnottib{D. Francia, A. Sagnotti, Phys. Lett. B53 (2002) 303}
\lref\sagnottic{D. Francia, A. Sagnotti, Class. Quant. Grav.
 20 (2003) S473}
\lref\sagnottid{D. Francia, J. Mourad, A. Sagnotti, Nucl. Phys. B773
(2007) 203}
\lref\labastidaa{ J. Labastida, Nucl. Phys. B322 (1989)}
\lref\labastidab{ J. Labastida, Phys. rev. Lett. 58 (1987) 632}
\lref\mvd{L. Brink, R.Metsaev, M. Vasiliev, Nucl. Phys. B 586 (2000)183}
\lref\klebanov{ I. Klebanov, A. M. Polyakov,
Phys.Lett.B550 (2002) 213-219}
\lref\mve{
X. Bekaert, S. Cnockaert, C. Iazeolla,
M.A. Vasiliev,  IHES-P-04-47, ULB-TH-04-26, ROM2F-04-29, 
FIAN-TD-17-04, Sep 2005 86pp.}
\lref\sagnottie{A. Campoleoni, D. Francia, J. Mourad, A.
 Sagnotti, Nucl. Phys. B815 (2009) 289-367}
\lref\sagnottif{
A. Campoleoni, D. Francia, J. Mourad, A.
 Sagnotti, arXiv:0904.4447}
\lref\sagnottig{D. Francia, A. Sagnotti, J.Phys.Conf.Ser.33:57 (2006)}
\lref\selfa{D. Polyakov, Int.J.Mod.Phys.A20:4001-4020,2005}
\lref\selfb{ D. Polyakov, arXiv:0905.4858}
\lref\selfc{D. Polyakov, arXiv:0906.3663}
\lref\selfd{D. Polyakov, Phys.Rev.D65:084041 (2002)}
\lref\mirian{A. Fotopoulos, M. Tsulaia, Phys.Rev.D76:025014,2007}
\lref\extraa{I. Buchbinder, V. Krykhtin,  arXiv:0707.2181}
\lref\extrab{I. Buchbinder, V. Krykhtin, Phys.Lett.B656:253-264,2007}
\lref\extrac{X. Bekaert, I. Buchbinder, A. Pashnev, M. Tsulaia,
Class.Quant.Grav. 21 (2004) S1457-1464}
\lref \extrad{I. Buchbinder, A. Pashnev, M. Tsulaia,
arXiv:hep-th/0109067}
\lref\extraf{I. Buchbinder, A. Pashnev, M. Tsulaia, 
Phys.Lett.B523:338-346,2001}
\lref\extrag{I. Buchbinder, E. Fradkin, S. Lyakhovich, V. Pershin,
Phys.Lett. B304 (1993) 239-248}
\lref\extrah{I. Buchbinder, A. Fotopoulos, A. Petkou, 
 Phys.Rev.D74:105018,2006}
\lref\bonellia{G. Bonelli, Nucl.Phys.B {669} (2003) 159}
\lref\bonellib{G. Bonelli, JHEP 0311 (2003) 028}
\lref\ouva{C. Aulakh, I. Koh, S. Ouvry, Phys. Lett. 173B (1986) 284}
\lref\ouvab{S. Ouvry, J. Stern, Phys. Lett.  177B (1986) 335}
\lref\ouvac{I. Koh, S. Ouvry, Phys. Lett.  179B (1986) 115 }
\lref\hsself{D.Polyakov, arXiv:1005.5512}
\lref\sundborg{ B. Sundborg, ucl.Phys.Proc.Suppl. 102 (2001)}
\lref\sezgin{E. Sezgin and P. Sundell,
Nucl.Phys.B644:303- 370,2002}
\lref\morales{M. Bianchi,
J.F. Morales and H. Samtleben, JHEP 0307 (2003) 062}
\lref\giombif{S. Giombi, Xi Yin, arXiv:0912.5105}
\lref\giombis{S. Giombi, Xi Yin, arXiv:1004.3736}
\lref\bekaert{X. Bekaert, N. Boulanger, P. Sundell, arXiv:1007.0435}
\lref\taronna{A. Sagnotti, M. Taronna, arXiv:1006.5242, 
Nucl.Phys.B842:299-361,2011}
\lref\zinoviev{Yu. Zinoviev, arXiv:1007.0158}
\lref\fotopoulos{A. Fotopoulos, M. Tsulaia, arXiv:1007.0747}
\lref\fotopouloss{A. Fotopoulos, M. Tsulaia, arXiv:1009.0727}
\lref\taronnao{M. Taronna, arXiv:1005.3061}
\lref\per{	
N. Boulanger,S. Leclercq, P. Sundell, JHEP 0808(2008) 056 }
\lref\mav{V. E. Lopatin and M. A. Vasiliev, Mod. Phys. Lett. A 3 (1988) 257}
\lref\zinov{Yu. Zinoviev, Nucl. Phys. B 808 (2009)}
\lref\sv{E.D. Skvortsov, M.A. Vasiliev,
Nucl. Phys.B 756 (2006)117}
\lref\mvasiliev{D.S. Ponomarev, M.A. Vasiliev, Nucl.Phys.B839:466-498,2010}
\lref\zhenya{E.D. Skvortsov, Yu.M. Zinoviev, arXiv:1007.4944}
\lref\perf{N. Boulanger, C. Iazeolla, P. Sundell, JHEP 0907 (2009) 013 }
\lref\pers{N. Boulanger, C. Iazeolla, P. Sundell, JHEP 0907 (2009) 014 }
\lref\selft{D. Polyakov,Phys.Rev.D82:066005,2010}
\lref\selftt{D. Polyakov, Int.J.Mod.Phys.A25:4623-4640,2010}
\lref\tseytlin{I. Klebanov, A Tseytlin, Nucl.Phys.B546:155-181,1999}
\lref\ruben{R. Manvelyan, K. Mkrtchyan, W. Ruehl, arXiv:1009.1054}
\lref\rubenf{R. Manvelyan, K. Mkrtchyan, W. Ruehl, Nucl.Phys.B836:204-221,2010}
\lref\xavierf{X. Bekaert, J. Erdmenger, D. Ponomarev, C. Sleight, arXiv: 1508.04292}
\lref\xaviers{X. Bekaert, J. Erdmenger, D. Ponomarev, C. Sleight, JHEP 1503 (2015) 170}
\lref\antalf{ A. Jevicki, K. Jin, Q. Ye, J.Phys. A46 (2013) 214005}
\lref\mtaronna{M. Taronna,  arXiv:1209.5755}
\lref\metsaevf{R. R. Metsaev, J.Phys. A46 (2013) 214021}
\lref\antals{ A. Jevicki, K. Jin, Q. Ye,  J.Phys. A44 (2011) 465402}
\lref\oli{O. Schlotterer,  Nucl.Phys. B849 (2011) 433-460}
\lref\bbfran{X. Bekaert, N. Boulanger, D. Francia, J.Phys. A48 (2015) 22, 225401 }
\lref\sftmartin{M. Schnabl, Adv.Theor.Math.Phys. 10 (2006) 433-501}
\lref\selfquartic{D. Polyakov, Phys.Rev. D83 (2011) 046005}
\lref\selfframe{D. Polyakov, Phys.Rev. D84 (2011) 126004}
\lref\selfframes{D. Polyakov, Phys.Rev. D89 (2014) 2, 026010}
\lref\selfsft{D. Polyakov,  arXiv:1507.06226 , to appear in Phys.Rev. D}

\centerline{\bf  1. Introduction}

Interacting higher-spin gauge fields are known to be the crucial ingredient
of AdS/CFT and holography in general and, at the same time, difficult and fascinating
objects to work with.
Despite the fact that the higher spin theories in $AdS$ spaces  can circumvent 
the restrictions imposed by the Coleman-Mandula's theorem, describing
the gauge-invariant higher-spin interactions is a highly nontrivial 
 problem since the gauge symmetry in these theories must be 
sufficiently powerful in order to eliminate unphysical degrees of freedom

~{\fronsdal,  \sagnottia, \sagnottib, \sagnottic,
\sagnottid, \sorokin, 
\mva, \mvb, \mvc, \mvasiliev, \deser, \bengt, \siegel, \siegelb,
\nicolai, \damour, \brink, \boulanger,
\labastidaa, \labastidab, \mvd, \mve, \mvasiliev, \mirian, \extrah,
\bonellia,\extrad, \bekaert, 
\perf, \pers, \taronna, \taronnao, 
\zinoviev,
\fotopoulos, \fotopouloss, \taronna,\zinoviev, \bekaert, \morales, \giombif, 
\giombis,
\sezgin, \sundborg, \per, \zhenya, \ruben, \rubenf}

The restrictions imposed by such a gauge symmetry make understanding the
interactions in higher-spin theories a distinctively complicated problem.
While there was some progress in classification of the higher-spin 3-vertices
and cubic interactions over recent years, our understandingof  the higher-order interactions is
still largely incomplete even at the quartic level (with the quartic interactions
presumably related to conformal blocks in the dual CFT's). 
Apart from that, at this point we know very little, if anything
 about higher spin theories beyond the quartic order,
e.g. about quintic and higher order interactions. One general property expected 
from the interacting higher spin theories, 
is that they have to be essentially nonlocal.
Such a nonlocality  is a natural compromise in order to evade the restrictions
imposed by Coleman-Mandula's theorem in flat space-time; however, it also appears to persist
in curved background geometries such as AdS where , at least formally, the Coleman-Mandula's restrictions
can be circumvented. 
In fact, this nonlocality property can be seen in a very natural way in string theory, where one interprets the
higher-spin fields as space-time wavefunctions for the certain class of vertex operators
and the higher spin interactions in terms of the worldsheet correlators of these operators.
Indeed, string theory appears to be a natural framework to describe higher spin dynamics,
as the frame-like description of the higher spin modes in Vasiliev's formalism
has a very natural vertex operator interpretation ~{\selfframe}.
The vertex operators describing higher spin modes have, however, the ghost structure
which is very different from that of operators for the lower spin modes, such as a photon 
{\selfframe, \selfframes}
First of all, these operators couple nontrivially to the $\beta-\gamma$ system
of superconformal ghosts. These couplings are classified by the minimum
superconformal ghost pictures carried by the operators. That is, the vertex operators
of spin $s$ are the elements of ghost cohomologies $H_{-s}\sim{H_{s-2}}(s>2)$
implying that they are annihilated by direct picture changing transformation at negative
picture $-s$ and by inverse picture changing transformation at $dual$ positive picture $s-2$.
Moreover, these operators also have an anomalous $b-c$ ghost number couplings, which is
closely related to the nonlocalities in the resulting interactions.
Namely, the standard $b-c$ pictures for lower spin vertices, such as a photon,
involve either integrated form at ghost number 0, or unintegrated form at ghost number 1.
 Combined with $b-c$ ghost number anomaly cancellation condition,
this requires 3 unintegrated operators and $N-3$ integrated,leading, 
for example,
to the standard form of Veneziano amplitude, defining $local$ quartic
interaction terms in the low-energy effective action.
The ghost structure of the higher-spin operators is different. Since 
these operators violate picture equivalence and have no picture 0
 representation, one needs operators both at positive and negative 
superconformal ghost pictures to ensure the correct superconformal ghost number
balance in correlators. The higher spin vertex operators at positive pictures,
however, only exist in the integrated form, given by the integrals
of three types of terms. That is, for spin $s$ operators these types
carry the ghost structures  $e^{(s-2)\phi},c{e^{(s-3)\phi+\chi}}$
and ${\partial{c}{c}e^{(s-4)\phi+2\chi}}$ respectively (where $\phi$ and $\chi$
are the bosonized superconformal ghosts).
Typically, it is the first and the second types of terms that contribute
to higher spin correlators. The second type, while being integrated
operators, also carry the $b-c$ ghost number one. This leads to the possibility
of the $b-c$ ghost number balance being saturated while having an extra
integration in the correlator (compared to the correlators of vertices for lower spins). 
This extra integration  leads to appearance of additional
singularities in the resulting scattering amplitude. Unlike the poles
of lower spin amplitudes (such as Veneziano amplitude), 
corresponding to particle exchanges, the extra singularities reflect the
nonlocalities in the related interaction terms
for higher spin modes in the low-energy effective
action. Typically, the combination of
superconformal and $b-c$  ghost number balance constraints dictates that
the type $2$ operators always contribute to the higher spin amplitudes,
leading to nonlocalities in space-time (e.g. considered in {\selfquartic}).
However, as we point out in this work, there exists a class of amplitudes
contributed by the first type operators only. These amplitudes are in turn 
related to the appearance of
local gauge-invariant interactions for higher spins in the low-energy effective action. 
We find that the higher spin interaction vertices of this type:

1)  only appear at higher orders of interaction, 
starting from the quintic order.
The interactions of this type are absent at lower orders, such as the
 quartic order.

2) only may appear if the spin values satisfy the localization constraint,that
is, for the order $N$ higher spin interaction the sum of 3 spins
must be roughly equal to the sum of the  remaining $N-3$, i.e.
\eqn\grav{\eqalign{s_1+s_2+s_3=s_4+...+s_N+\alpha}}
where $\alpha$ is of the order of 1 (in the concrete example of the quintic
interaction, considered in this work, $\alpha=2$).
The rest of this paper is organized as follows.
In the next section, we review the vertex operator formalism for
frame-like higher spin gauge fields of arbitrary spin values and 
perform a general analysis of derivative structure of higher spin interaction,
related to the correlators of these operators.
In the Section 3, we calculate the particular example of the 
5-point amplitude , satisfying
the constraint (1), leading to the localization effect in the quintics.
The structure of the amplitude and of the interaction particularly simplifies
if one of the spins, $s_4$ or $s_5$ has a relatively small value,
up to 6 (in this work we consider the case $s_4=4$, with all other spin values
arbitrary, up to localization constraint (1)).
In the calculation, we particularly use the OPE formalism for the Bell 
polynomial operators , developed in the previous work ~{\selfsft}
In the concluding section, we discuss the physical implications of the 
calculation done in this work and its possible generalizations.

\centerline{\bf 2. Frame-like Fields and Vertex Operators:}

\centerline{\bf 
Review of the Formalism and Preliminary Derivative Analysis}

In the frame-like formalism a symmetric 
higher spin field of spin $s$ is described
by the set of of $s$ two-row gauge fields
$\Omega^{s-1|t}(x)\equiv\Omega_m^{a_1...a_{s-1}|b_1...b_t}(x)(0\leq{t}\leq{s-1})$
appearing in the higher spin extension of Cartan's 1-form by
generators of infinite-dimensional higher spin algebra:
\eqn\grav{\eqalign{
\Omega^{(1)}\equiv\Omega_m{dx^m}=(
e_m^a{T_a}+\omega_m^{a_1a_2}T_{a_1a_2}
+\sum_{s=3}^\infty\sum_{t=0}^{s-1}\Omega_m^{a_1...a_{s-1}|b_1...b_t}T_{a_1...a_{s-1}|b_1...b_t})dx^m
}}
where $T_a,T_{a_1a_2}$ are the space-time isometry generators,
$e$ and $\omega$ are the vielbeins and spin connections for $s=2$
 and
$T_{a_1...a_{s-1}|b_1...b_t}$ are the generators of the higher-spin algebra that envelops
the space-time isometry algebra.
In this formalism, only $t=0$ fields are genuinely dynamical 
and are related to Fronsdal's metric-type
higher spin fields upon symmetrization.
The fields with $t\neq{0}$ are the $extra$ fields, 
related to  the Fronsdal's field $\Omega^{s-1|0}$
by generalized  zero torsion constraints
according to
\eqn\grav{\eqalign{\Omega^{s-1|t}(x)\sim\partial^t\Omega^{s-1|0}}}
In curved backgrounds, such as AdS, the frame-like approach to higher spin dynamics 
is remarkably efficient.
At the same time,  string theory is known to be a particularly natural framework
to  describe the interactions of higher spin fields in terms of the worldsheet
correlators of the vertex operators.
 In particular, the bosonic string spectrum in the tensionless
limit contains vertex operators that can be interpreted as sourses for Fronsdal's 
higher spin modes, although the correlators calculated in this  limit are somewhat difficult
to interpret in terms of higher spin interactions in the field-theoretic low-energy limit
of string theory ~{\sagnottia, \sagnottib, 
\sagnottic, \sagnottid, \taronna, \taronnao}. 
Apart from the tensionless limit, the vertex operators 
describing massless higher spin modes can be constructed
 in RNS superstring theory.
These operators are the elements of nonzero ghost cohomologies $H_{-s}\sim{H_{s-2}}$,
existing either at minimal negative picture $s$ and below, or dual minimum positive picture
$s-2$ and above. These operators describe massless fields
 of spin $s\geq{3}$ in the $frame-like$
formalism (except for $s=3$ where the corresponding
 higher spin mode is actually a Fronsdal's field).
BRST invariance constraints on these operators entail the on-shell constraints on the 
higher spin fields (such as Pauli-Fierz constraints) while BRST nontriviality conditions
entail the gauge and diffeomorphism transformations. As these transformations shift the
vertex operators by BRST-exact terms, the resulting correlation functions are gauge-invariant by
 construction.
For $s=3$,the manifest expressions for the vertex operators in
RNS superstring theory  are rather 
simple and are given by ~{\selft, \selfframe, \selfframes}
\eqn\grav{\eqalign{V_{2|0}(p,z)=\Omega_{m}^{ab}(p)ce^{-3\phi}
\psi^m\partial{X_a}\partial{X_b}e^{ipX}(z)}}
at unintegrated minimal negative ghost picture $-3$
and
\eqn\grav{\eqalign{V_{2|0}(p,z)
=\Omega_{m}^{ab}(p)K\circ\int{dz}e^{\phi}\psi^m\partial{X_a}\partial{X_b}e^{ipX}(z)}}
at minimal positive picture $+1$
which is always integrated.
Here $X^m$ are the target space coordinates, $\psi^m$ are the RNS fermions
 and $K$ is the homotopy transform , necessary to ensure the
BRST invariance of the positive picture operator (5).
 (see ~{\selft, \selfframe, \selfframes}
 for the detailed description
of the transform). 

The operators (4), (5) are thus the elements of $H_{-3}\circ{H_1}$. 

The manifest expressions for the vertex operators $V_{s-1|t}$ 
for the frame-like fields with $s>3$ 
become far more complicated, however, a significant simplification occurs for the case
$t=s-3$. In this case, the explicit expressions for the vertex operators
are given by ~{\selfframes}
\eqn\grav{\eqalign{V_{s-1|s-3}^{(-)}=ce^{-s\phi}\partial{X_{m_1}}...\partial{X_{s-1}}\psi^{\alpha_0}
\partial\psi_{\alpha_1}\partial^2\psi_{\alpha_2}...\partial^{s-3}\psi_{\alpha_{s-3}}e^{ipX}
\Omega_{\alpha_0}^{m_1...m_{s-1}|\alpha_1...\alpha_{s-3}}(p)}}
at the minimal negative unintegrated picture
and
\eqn\grav{\eqalign{V_{s-1|s-3}^{(+)}=
K\circ{\int}{dz}e^{s-2}\partial{X_{m_1}}...\partial{X_{s-1}}
\psi^{\alpha_0}\partial\psi_{\alpha_1}\partial^2\psi_{\alpha_2}...\partial^{s-3}\psi_{\alpha_{s-3}}e^{ipX}
\cr\times
\Omega_{\alpha_0}^{m_1...m_{s-1}|\alpha_1...\alpha_{s-3}}(p)
}}
at the minimal positive picture representation
with the result of the homotopy K-transformation given explicitly by

\eqn\grav{\eqalign{
K\circ{\oint}{dz}e^{s-2}\partial{X_{m_1}}...\partial{X_{m_{s-1}}}
\psi^{\alpha_0}\partial\psi_{\alpha_1}\partial^2\psi_{\alpha_2}
...\partial^{s-3}\psi_{\alpha_{s-3}}e^{ipX}
\Omega_{\alpha_0}^{m_1...m_{s-1}|\alpha_1...\alpha_{s-3}}(p)
\cr
={A_0(p;u)+A_1(p;u)+A_2(p;u)}}}
where
\eqn\grav{\eqalign{
A_0(p;u)=
\Omega_{\alpha_0}^{m_1...m_{s-1}|\alpha_1...\alpha_{s-3}}(p)
\cr\times
\oint{dz}(z-u)^{2s-4}B^{(2s-4)}_{2\phi-2\chi-\sigma}
e^{(s-2)\phi}\psi^{\alpha_0}\partial\psi_{\alpha_1}\partial^2\psi_{\alpha_2}
...\partial^{s-3}\psi_{\alpha_{s-3}}e^{ipX}(u)\cr
A_1(p;u)=2\Omega_{\alpha_0}^{m_1...m_{s-1}|\alpha_1...\alpha_{s-3}}(p)
\oint{dz}(z-u)^{2s-4}{c}e^{\chi+(s-3)\phi+ipX}\cr\times
\lbrace\sum_{k=0}^{s-3}
(-1)^{k+1}k!\lbrack
\sum_{j=0}^{s-2+k}B^{(s+k-1)}_{\phi-\chi}\partial{X_{m_1}}...\partial{X_{m_{s-1}}}
\psi^{\alpha_0}
\cr\times
\partial\psi_{\alpha_1}...\partial^{k-1}\psi_{\alpha_{k-1}}
(-ip_{\alpha_k})\partial^{k+1}\psi_{\alpha_{k+1}}...
\partial^{s-3}\psi_{\alpha_{s-3}}
\cr+
\sum_{j}^{s-2+k}{1\over{j!}}B^{(s-2+k-j)}_{\phi-\chi}
\partial{X_{m_1}}...\partial{X_{m_{s-1}}}
\cr\times
\psi^{\alpha_0}\partial\psi_{\alpha_1}
...\partial^{k-1}\psi_{\alpha_{k-1}}
(\partial^{1+j}X_{\alpha_k})\partial^{k+1}\psi_{\alpha_{k+1}}...
\partial^{s-3}\psi_{\alpha_{s-3}}\rbrack
\cr
-2(s-1)\sum_{k=0}^{s-1}{1\over{k!}}B^{(s-1-k)}_{\phi-\chi}
\partial{X_{m_1}}...\partial{X_{m_{s-2}}}\partial^j\psi_{m_{s-1}}
\psi^{\alpha_0}\partial\psi_{\alpha_1}\partial^2\psi_{\alpha_2}
...\partial^{s-3}\psi_{\alpha_{s-3}}
\cr
A_2(p;u)=-4(2s-3)
\Omega_{\alpha_0}^{m_1...m_{s-1}|\alpha_1...\alpha_{s-3}}(p)
\cr\times
\oint{dz}(z-u)^{2s-4}\partial{c}c
e^{2\chi+(s-4)\phi}\psi^{\alpha_0}\partial\psi_{\alpha_1}\partial^2\psi_{\alpha_2}
...\partial^{s-3}\psi_{\alpha_{s-3}}e^{ipX}(u)
}}
with $A_0$, $A_1$ and $A_2$ terms having the ghost structure described above, namely,
 $e^{(s-2)\phi},c{e^{(s-3)\phi+\chi}}$
and ${\partial{c}{c}e^{(s-4)\phi+2\chi}}$  respectively 
and the integrals  being taken around an arbitrary point $u$ on the worldsheet.
The choice of $u$ is arbitrary as any 
correlators involving  ${V_{s-1|s-3}^{(+)}(p;u)}$-operators
are $u$-independent since their $u$-derivatives are BRST-exact in the
small Hilbert space; the operators themselves can be cast as BRST 
commutators in the large
Hilbert space (but not in the small Hilbert space)
and therefore are the elements of BRST cohomology in the small Hilbert space
,given the on-shell conditions on $\Omega$ and 
modulo the gauge transformations ~{\selft, \selfframes}.
The $B^{(n)}_{m\phi+n\chi+p\sigma}$  are the normalized degree $n$ Bell polynomials
in the bosonized superconformal ghost fields $\phi$, $\chi$ and $\sigma$,
defined according to
\eqn\grav{\eqalign{\partial_z^{n}e^{m\phi+n\chi+p\sigma}(z)\equiv
{1\over{n!}}:B^{(n)}_{m\phi+n\chi+p\sigma}e^{m\phi+n\chi+p\sigma}(z)}}
where $m,n,p$ are some numbers. The properties and the operator algebras
involving the Bell polynomial operators have been studied in details
in ~{\selfsft} and will be used in the present calculation.
Although the negative and positive picture representations
of the higher spin vertex operators
 ${V_{s-1|s-3}^{(\pm)}(p;u)}$ are not directly related 
by picture-changing transformations (as is clear from the fact that they
belong to the nonzero negative/positive cohomologies), they can be mapped
to each other by combining BRST-invariant picture-changing and 
$Z$-transformations which ensures that the on-shell BRST-invariance constraints
on $\Omega$ are identical in the both of the representations {\selfframe, \selfframes}.
Technically, it is easier to analyze these constraints  at negative
pictures, with the BRST invariance imposing the on-shell conditions:
\eqn\grav{\eqalign{
p^2\Omega_{\alpha_0}^{m_1...m_{s-1}|\alpha_1...\alpha_{s-3}}(p)=0\cr
\Omega_{\alpha_0m}^{mm_3...m_{s-1}|\alpha_1...\alpha_{s-3}}(p)=0\cr
\Omega_{\alpha_0\alpha}^{m_1...m_{s-1}|\alpha\alpha_3...\alpha_{s-3}}(p)=0\cr
\Omega_{\alpha_0\alpha}^{m_2...m_{s-1}|\alpha\alpha_2...\alpha_{s-3}}(p)=0}}
The first and the second constraints  are the standard  constraints for
the symmetric higher-spin fields in the frame-like description.
They are supplemented by two more constraints, indicating that the vertex 
operators (6), (7)
 describe the frame-like fields in space-time with the gauge
partially fixed. 
The gauge transformations for $\Omega$:
\eqn\grav{\eqalign{
\Omega_m^{a_1...a_{s-1}|b_1...b_{s-3}}\rightarrow
\Omega_m^{a_1...a_{s-1}|b_1...b_{s-3}}+p_m\Lambda^{a_1...a_{s-1}|b_1...b_{s-3}}}}
in turn shift the operators (6), (7) by the BRST-exact term 
(see ~{\selfframe, \selfframes} for the detailed BRST analysis).
It is essential that the $\beta$-function equations for the $\Omega$-field,
in the leading order, defining the kinetic term for the corresponding
Fronsdal's field and
obtained using the $off-shell$ Weyl invariance constraints on the 
operators (6), (7) differ from the on-shell BRST-invariance constraints 
due to nontrivial ghost couplings of the higher spin vertex operators.
Namely, the leading order $\beta_\Omega=0$ equations are equivalent
to ${\hat{L}}_{AdS}\Omega=0$ where ${\hat{L}}_{AdS}$ is the Fronsdal's kinetic
operator in  the $AdS$ , rather than flat space, as was shown
by detailed computations in ~{\selfframes}. In particular, the appearance
of the mass-like term in  ${\hat{L}}_{AdS}$ (which is identified 
with the tail of the covariant derivatives in the AdS Laplacian)
is directly related to the Weyl transformations of the anomalous 
superconformal ghost parts of the higher spin vertex operators.

For the lower spin vertex operators,
such as a graviton, there is a familiar example, somewhat
reminiscent of this difference between Weyl and BRST transformations:
e.g. recall that the graviton's $\beta$-function contains a term 
given by the second derivative of the dilaton (which is of course absent
in the on-shell conditions imposed by the BRST-invariance).
For this reason , it is natural to think of the $(m,\alpha)$
indices in (11) as of those living in the tangent bundle of 
emergent $AdS$ space. Accordingly, the correlators of these operators
describe the $AdS$ higher spin interactions, upon the pullback from the
bundle to the manifold.

For the $t$ values other than $t=s-3$ the manifest expressions for the
higher spin vertex operators become significantly more complicated.

The explicit relation between the vertex operators 
$V_{s-1|t}\equiv{\Omega^{s-1|t}{W_{s-1|t}}}$ 
(with $W$ being conformal dimension 0 primary fields
which, upon coupling with the space-time higher spin frame-like 
fields $\Omega^{s-1|t}$
are the elements of $H_{-s}$)
with different
$t$ values , generating the chain of the zero torsion constraints,
is given by (modulo the on-shell constraints (11))

\eqn\grav{\eqalign{ \Omega^{s-1|t}:\Gamma{W_{s-1|t}}:
=\Omega^{s-1|t+1}{W_{s-1|t+1}}}}

Here $W$ are the conformal dimension 0 vertex operators
which, upon coupling to the space-time higher spin frame-like 
fields $\Omega^{s-1|t}$, become 
 the elements of $H_{-s}$.
$\Gamma$ is the picture-changing operator for the $\beta-\gamma$
ghost system given by
\eqn\grav{\eqalign{\Gamma=\lbrace{Q},e^\chi\rbrace=
-{1\over2}e^\phi\psi_m\partial{X^m}+{1\over4}be^{2\phi-\chi}
(\partial_\chi+\partial\sigma)+ce^\chi\partial\chi}}

The relation (13) can be generalized according to
\eqn\grav{\eqalign{ \Omega^{s-1|t}:\Gamma^k{W_{s-1|t}}:
=\Omega^{s-1|t+1}{W_{s-1|t+k}}(k\leq{s-3-t})\cr
 \Omega^{s-1|t}:\Gamma^k{W_{s-1|t}}:=0(k>{s-3-t})
}}
where $\Gamma^k=:\Gamma....\Gamma:=:e^{k\phi}G\partial{G}...\partial^{k-1}G:$
is the normally ordered product of $k$ picture-changing operators
and $G$ is the full matter$+$ghost worldsheet supercurrent.
In particular,
to obtain the operator for the Fronsdal's field one can take $k=s-3$
and $t=0$.

$k=s-3$
Note that all the $V_{s-1|t}\equiv{\Omega^{s-1|t}{W_{s-1|t}}}$ higher spin
vertex operators are the elements of $H_{-s}\sim{H_{s-2}}$ cohomology
for all the values of $0\leq{t}\leq{s-3}$, although the canonical pictures for
$V_{s-1|t}$ are different and equal to
 $-2s+t+3$ at the negative picture representation.
The action of $\Gamma$ on $V_{s|t}$ thus results in increasing the ghost picture 
by one unit and the appearance of the extra $p$ factor in front of $\Omega(p)$,
typically
due to the contraction of the first term in $\Gamma$ with $e^{ipX}$ factor
in the vertex operator.
The zero torsion constraints can of course be reformulated
equivalently for the operators in the positive cohomology representations;
will all the spin $s$ operators being the elements of $H_{s-2}$,
the constraints for the frame-like vertex operators in the positive 
cohomologies are
\eqn\grav{\eqalign{ \Omega^{s-1|t}:\Gamma^{-k}{W_{s-1|t}}:
=\Omega^{s-1|t+1}{W_{s-1|t+k}}(k\leq{s-3-t})\cr
 \Omega^{s-1|t}:\Gamma^{-k}{W_{s-1|t}}:=0(k>{s-3-t})
}}
with $ \Omega^{s-1|t}W_{s-1|t}$-operators having canonical
ghost pictures $2s-5-t$, in particular the operator for the Fronsdal's
field having canonical positive picture $2s-5$.
Here $:\Gamma^{-k}:=:(\Gamma^{-1})^k:$ with 
$\Gamma^{-1}=-4ce^{\chi-2\phi}\partial\chi$ being the $inverse$ picture-changing
operator.

In practice, the equations (15), (16) generating the generalized 
zero torsion constraints by the picture-changing relations, 
are hard to solve.
For example, the solution for the simplest of the equations (15) 
for $t=s-4$ is
already complicated enough, with the vertex operator having the form
\eqn\grav{\eqalign{
V_{s-1|s-4}(p;z){\sim}ce^{-(s+1)\phi}\partial{X_{m_1}}...\partial{X_{s-1}}\psi^{\alpha_0}
\partial\psi_{\alpha_1}\partial^2\psi_{\alpha_2}...\partial^{s-4}\psi_{\alpha_{s-4}}
\cr\times
\lbrack
\alpha{B^{(2s-3)}_{-\phi}}
\sum_{p=0}^{s-3}\sum_{q=p+1}^{s-2}
\alpha_{p|q}\partial^p\psi_m\partial^q\psi^nB^{(2s-4-p-q)}_{-\phi}
+\alpha_{s-3|s-1}\partial^{s-3}\psi_m\partial^{s-1}\psi^m\rbrack
e^{ipX}}}
at canonical $-s-1$-picture 
where the $\alpha_{p|q}$-coefficients must be calculated
so as to ensure
that $V_{s-1|s-4}(p;z)$ is primary (i.e. the singularities
of cubic and higher orders stemming from the OPE of the stress-tensor
with the $\psi$-part must be cancelled by those stemming from the operator
product with the Bell polynomials in the derivatives
of the  $\phi$-ghost field). Note that 
the leading order  of the operator products of all the Bell polynomials
$B_{-\phi}^{(N)}$ with $e^\phi$ is the simple pole for any $N$, which ensures
that the picture-changing transformation of (17) does not produce any terms other than those proportional to $V_{s-1|s-3}$, as well as the absence of singularities
in  the OPE of $\Gamma$ and  $V_{s-1|s-4}(p;z)$. It is now not hard to see that
the expressions for operators  with $t\leq{s-5}$ will get more and more tedious
for the lower $t$ values; their general structure would involve
sums over
multiple products of $\partial^p{\vec{\psi}}\partial^q{\vec{\psi}}$-factors
multiplied by products of $s-3-t$ Bell polynomials of the ghost fields,
making them cumbersome objects to work with.

Thus the $V_{s-1|s-3}$-operators appear
to be  particularly convenient and natural objects to use in order
to describe the
higher-spin vertices in space-time.
 However, there are
  restrictions
on the spin values for the vertices that can be described 
by the correlators with all the operators being of the
$V_{s-1|s-3}$-only. That is, for a $n=p+q$-point 
higher spin amplitude containing
$p$ $V_{s-1|s-3}$-operators at positive cohomologies and $q$
operators at negative cohomologies the spin values must satisfy
\eqn\grav{\eqalign{\sum_{j=1}^{p}(s_j-2)-\sum_{j=p+1}^{p+q}s_j=-2}}
in order to  satisfy the superconformal ghost balance constraint.
The amplitudes with spin values not satisfying (18) cannot be described by
using solely the  operators of this type and require finding explicit
solutions of the operator
equations (15), (16) making them far more complicated.
At the same time, note that if the constraint (18) is satisfied,
despite the fact that the correlation functions of
the $V_{s-1|s-3}$-type operators by construction
contain certain minimal number of derivatives
(since each space-time field $\Omega_{s-1|s-3}$
 by definition contains $s-3$ derivatives), all the amplitudes 
involving $V_{s-1|s-3}$  can be cast equivalently
 in terms of those involving
operators for the Fronsdal's fields and/or the operators for
the extra fields $\Omega_{s-1|t}$ with lower $t$, using the zero
torsion relations (15), (16), combined with the picture equivalence of the
operators inside each particular cohomology.
For example, consider a 5-point higher spin amplitude with 
the spins satisfying the constraints (18): $s_1+s_2-s_3-s_4-s_5=2$.
In the amplitude of this type, two operators are
integrated at positive picture, and
three operators are unintegrated at negative pictures.
Using the zero torsion relations (15), (16)
we get
\eqn\grav{\eqalign{A(s_1,...s_5)=
<\Omega^{s_1-1|s_1-3}W_{s_1-1|s_1-3}^{(s_1-2)}(p_1)
\Omega^{s_2-1|s_2-3}W_{s_2-1s_2-3}^{(s_2-2)}(p_2)
\cr
\Omega^{s_3-1|s_3-3}W_{s_3-1|s_3-3}^{(-s_3)}(p_3)
\Omega^{s_4-1|s_4-3}W_{s_4-1|s_4-3}^{(-s_4)}(p_4)
\Omega^{s_5-1|s_5-3}W_{s_5-1|s_5-3}^{(-s_5)}(p_5)>
\cr
=<:\Gamma^{-s_1-s_2+6}:
\Omega^{s_1-1|0}W_{s_1-1|s_1-3}^{(2s_1-5)}(p_1)
\Omega^{s_2-1|0}W_{s_2-1s_2-3}^{(2s_2-5)}(p_2)
\cr
:\Gamma^{s_3+s_4+s_5-9}:
\Omega^{s_3-1|0}W_{s_3-1|0}^{(3-2s_3)}(p_3)
\Omega^{s_4-1|0}W_{s_4-1|0}^{(3-2s_4)}(p_4)
\Omega^{s_5-1|0}W_{s_5-1|0}^{(3-2s_5)}(p_5)>
\cr=
<
\Omega^{s_1-1|0}W_{s_1-1|0}^{(2s_1-6)}(p_1)
\Omega^{s_2-1|0}W_{s_2-1|0}^{(2s_2-6)}(p_2)
\Omega^{s_3-1|0}W_{s_3-1|0}^{(2-2s_3)}(p_3)
\cr
\Omega^{s_4-1|0}W_{s_4-1|0}^{(2-2s_4)}(p_4)
\Omega^{s_5-1|0}W_{s_5-1|0}^{(2-2s_5)}(p_5)>
}}
i.e. the amplitude involving five $V_{s-1|s-3}$ higher spin operators
is identical to the one involving five Fronsdal operators
at pictures each lowered by one unit with respect to the canonical.

As it is clear from the above discussion,
the overall $b-c$ and superconformal ghost structures
of the $n=p+q$-point higher spin amplitudes, 
combined with the ghost structure
of the picture-changing operators (14) suggest the existence 
of two types of higher spin interactions at higher orders:
the first type having $p=n-3,q=3$. This amplitude 
involves 3 unintegrated negative picture  operators and
standard $N-3$ integrated operators at positive pictures.
All the terms, contributed by all the integrated operators, 
are of $A_0$-type (having $b-c$ ghost number zero)
The higher spin amplitudes of this type have the standard Veneziano pole
structure, leading to local interaction terms in the low-energy effective 
action (with the poles in the amplitudes corresponding to different
channels of particle exchanges).
These poles do not produce any physical nonlocalities 
(the nonlocalities that one may encounter upon the  space-time 
momentum integration and expressing the low-energy effective action in the
position space, are not physical and can be removed by substituting
the $\beta$-function equations at lower orders ~{\selfquartic}
(strictly speaking, the locality of the amplitude does not by itself guarantee the locality
of the related higher spin interaction vertex in the low-energy effective action;
for the case, considered in this paper see, however, the discussion below in the section 4)

The amplitudes of the second type, on the other hand,
have the structure $p=n-2,q=2$.They involve two unintegrated operators
contributing two $c$-ghosts, with the third $c$-ghost stemming from
the $A_1$-type terms of one of the integrated operators and the remaining
operators contributing $A_0$-type terms.
These amplitudes have the structure very different from those of the first 
type. Due to the extra integration, they contain extra poles, corresponding
to physical nonlocalities, rather than particle exchanges.
Unlike the Veneziano-type case, the nonlocalities in the position
space , obtained upon
the Fourier transform, cannot be removed using the $\beta$-function flows,
but reflect genuine  nonlocalities of the higher spin interactions.

In fact, generically most of the higher spin amplitudes are of the second type,
with the first type emerging only for the special combination of the spin 
values. We will refer to the appearance of the first type higher spin amplitudes
as the ``localization''.
In fact, as we shall point out below, such a localization 
effect does not occur at the quartic order but only appears at
quintic and higher order interactions.
In the next section, we will study this effect
by direct computation of the correlation functions.

\centerline {\bf 3. Quintic Interactions and Localization}

We start with the 4-point higher spin amplitude with the
spin values satisfying 
constraint (18) with $p=1,q=3$.
It is not difficult to see that this 4-point amplitude vanishes.
Indeed, the $V_{s_1}$ operator at positive cohomology
contains $s_1-2=s_1+s_2+s_3-2$ $\psi$-fields, which cannot
fully contract to the RNS fermions of the remaining 3 operators,
since the operators at negative pictures contribute altogether
$s_1+s_2+s_3-6$ $\psi$-fields.
This means that the $4$-point amplitude with such spin values
 admits no localization.
Next, consider the 5-point amplitude with the
localization constraint $p=2,q=3$.
To simplify the calculations as much as possible,
we assume that the value one of the spins
is small enough (namely, $s_4\equiv{u}=4$) and
$s_5+u-s_1-s_2-s_3=2$.
All the vertex operators for the frame-like fields are
in the $V_{s-1|s-3}$ representation, with
the operators the $s_1,s_2,s_3$ being unintegrated 
at negative pictures
and the operators for spins $u=4$ and $s_5=\sum{s}-2$
at integrated positive (to abbreviate the notations, 
denote $\sum{s}=s_1+s_2+s_3$). With such a picture arrangement,
only $A_0$-type terms of the both of the integrated operators
contribute to the correlator.
We are now all set to compute, step by step, the $X$,ghost and
$\psi$-factors of the correlator defining the quintic interaction
of the above spin values.
We start with the calculation of the $X$-part.
The correlation function is given by
\eqn\grav{\eqalign{A_X(p_1,...p_5|w;z;\xi)
=<\partial{X^{m_1}}...\partial{X^{m_{s_1-1}}}e^{ip_5{X}}(w)|_{w\rightarrow\infty}
\partial{X^{n_1}}...\partial{X^{n_{s_2-1}}}e^{ip_4{X}}(1)
\cr
\partial{X^{q_1}}...\partial{X^{q_{u-1}}}e^{ip_3{X}}(z)
\partial{X^{r_1}}...\partial{X^{r_{\sum{s}+1-u}}}e^{ip_2{X}}(\xi)
\partial{X^{t_1}}...\partial{X^{t_{s_3-1}}}e^{ip_1{X}}(0)>}}

where we have arranged the operator's insertions
in the order $(z_1=w)>(z_2=1)>z>\xi>(z_3=0)$, with
$z_{1,2,3}$ being the locations of the spin $s_{1,2,3}$ unintegrated 
vertices, $z$ and $\xi$ are the insertion points of the remaining spins
(to be integrated over). We shall later set $z_3\equiv{w}\rightarrow\infty$
but for now shall the $w$-dependence manifest to keep track
of the infinities.
To compute the correlator (20), define the partitions
\eqn\grav{\eqalign{
s_i-1=\sum_{j=1,j\neq{i}}^5(R_{ij}+Q_{ij});
i,j=1,...,5}}
where $R_{ij}$ is the number of contractions  between
$\partial{X}$'s of the operators of spins $s_{i}$ and $s_j$
(obviously $R_{ij}=R_{ji}$) and $Q_{ij}$ counts the contractions
of $X$-derivatives
of the operators for $s_i$ with the exponent $e^{ipX}$
in the operator for spin $s_j$.
Straightforward computation then gives:
\eqn\grav{\eqalign{
A_X(p_1,...p_5|w;z;\xi)
\cr
=\sum_{{{\lbrack}partitions:}s_i-1|\sum_{j=1,j\neq{i}}^5(R_{ij}+Q_{ij});i=1,...,5\rbrack}
{{6(s_1-1)!(s_2-1)!(s_3-1)!(\sum{s}-3)!}\over{
{\prod_{i=1}^4\prod_{j=2;j>i}^5R_{ij}!\prod_{k=1}^5\prod_{l=1;k{\neq}l}^5
Q_{kl}!}}} 
\cr
(i)^{\sum_{i,j=1;i\neq{j}}^5Q_{ij}}
(-1)^{\sum_{j=1}^5(Q_{1j}+Q_{2j})-Q_{21}+Q_{43}+Q_{43}+Q_{45}+Q_{53}}
w^{-\sum_{j=2}^5({2R_{1j}+Q_{1j}+Q_{1j}})}
\cr\times
(1-z)^{p_3p_4-2R_{24}-Q_{24}-Q_{42}}(1-\xi)^{p_2p_4-2R_{25}-Q_{25}-Q_{52}}
\cr
(z-\xi)^{p_2p_3-2R_{45}-Q_{45}-Q_{54}}z^{p_1p_3-2R_{34}-Q_{34}-Q_{43}}
\xi^{p_1p_2-2R_{35}-Q_{35}-Q_{53}}
\cr\times
p_1^{m_{Q_{12}+Q_{14}+Q_{15}+1}}...p_1^{m_{Q_{12}+Q_{14}+Q_{15}+Q_{13}}}
p_1^{n_{Q_{21}+Q_{24}+Q_{25}+1}}...p_1^{n_{Q_{21}+Q_{24}+Q_{25}+Q_{23}}}
\cr
p_1^{q_{Q_{41}+Q_{42}+Q_{45}+1}}...p_1^{q_{Q_{41}+Q_{42}+Q_{45}+Q_{43}}}
p_1^{r_{Q_{51}+Q_{52}+Q_{54}+1}}...p_1^{r_{Q_{51}+Q_{52}+Q_{54}+Q_{53}}}
\cr
p_2^{m_{Q_{12}+Q_{14}+1}}...p_2^{m_{Q_{12}+Q_{14}+Q_{15}}}
p_2^{n_{Q_{21}+Q_{24}+1}}...p_2^{n_{Q_{21}+Q_{24}+Q_{25}}}
\cr
p_2^{q_{Q_{41}+Q_{42}+1}}...p_2^{q_{Q_{41}+Q_{42}+Q_{45}}}
p_2^{t_{Q_{31}+Q_{32}+Q_{34}+1}}...p_2^{t_{Q_{31}+Q_{32}+Q_{34}+Q_{35}}}
\cr
{\times}p_3^{m_{Q_{12}+1}}...p_3^{m_{Q_{12}+Q_{14}}}
p_3^{n_{Q_{21}+1}}...p_3^{n_{Q_{21}+Q_{24}}}
\cr
p_3^{r_{Q_{51}+Q_{52}+1}}...p_3^{r_{Q_{51}+Q_{52}+Q_{54}}}
p_3^{t_{Q_{31}+Q_{32}+1}}...p_3^{t_{Q_{31}+Q_{32}+Q_{34}}}
\cr
p_4^{m_{1}}...p_4^{m_{Q_{12}}}
p_4^{q_{Q_{41}+1}}...p_4^{q_{Q_{41}+Q_{42}}}
p_4^{r_{Q_{51}+1}}...p_4^{r_{Q_{51}+Q_{52}}}
p_4^{t_{Q_{31}+1}}...p_4^{t_{Q_{31}+Q_{32}}}
\cr
p_5^{n_{1}}...p_5^{n_{Q_{21}}}
p_5^{q_{1}}...p_5^{q_{Q_{41}}}
p_5^{r_{1}}...p_5^{r_{Q_{51}}}
p_5^{t_{1}}...p_5^{t_{Q_{31}}}
\cr\times
\eta^{m_{\sum_{j}Q_{1j}+1}|n_{\sum_j{Q_{2j}+1}}}...\eta^{m_{\sum_{j}Q_{1j}+R_{12}}|n_{\sum_j{Q_{2j}+R_{12}}}}
\cr
\eta^{m_{\sum_{j}Q_{1j}+1+R_{12}}|q_{\sum_j{Q_{4j}+1}}}...\eta^{m_{\sum_{j}Q_{1j}+R_{12}+R_{14}}|q_{\sum_j{Q_{4j}+R_{14}}}}
\cr
\eta^{m_{\sum_{j}Q_{1j}+R_{12}+R_{14}+1}|r_{\sum_j{Q_{5j}+1}}}...\eta^{m_{\sum_{j}Q_{1j}+R_{12}+R_{14}+R_{15}}|r_{\sum_j{Q_{5j}+R_{15}}}}
\cr
\eta^{m_{\sum_{j}Q_{1j}+R_{12}+R_{14}+R_{15}+1}|t_{\sum_j{Q_{3j}+1}}}...\eta^{m_{\sum_{j}Q_{1j}+R_{12}+R_{14}+R_{15}+R_{13}}|t_{\sum_j{Q_{3j}+R_{13}}}}
\cr
\eta^{n_{\sum_{j}Q_{2j}+1+R_{12}}|q_{\sum_j{Q_{4j}+R_{14}+1}}}...\eta^{m_{\sum_{j}Q_{2j}+R_{12}+R_{24}}|q_{\sum_j{Q_{4j}+R_{14}}+R_{24}}}
\cr
\eta^{n_{\sum_{j}Q_{2j}+1+R_{12}+R_{24}}|r_{\sum_j{Q_{5j}+R_{15}+1}}}...\eta^{m_{\sum_{j}Q_{2j}+R_{12}+R_{24}+R_{25}}|r_{\sum_j{Q_{5j}+R_{15}}+R_{25}}}
\cr
\eta^{n_{\sum_{j}Q_{2j}+1+R_{12}+R_{24}+R_{25}+1}|t_{\sum_j{Q_{3j}+R_{31}+1}}}...\eta^{m_{\sum_{j}Q_{2j}+R_{12}+R_{24}+R_{25}}|t_{\sum_j{Q_{3j}+R_{31}}+R_{32}}}
\cr
\eta^{q_{\sum_{j}Q_{4j}+1+R_{41}+R_{42}}|r_{\sum_j{Q_{5j}+R_{51}+R_{52}+1}}}...\eta^{q_{\sum_{j}Q_{4j}+R_{41}+R_{42}+R_{45}}|r_{\sum_j{Q_{5j}+R_{15}}+R_{25}+R_{54}}}
\cr
\eta^{q_{\sum_{j}Q_{4j}+1+R_{41}+R_{42}+R_{45}}|t_{\sum_j{Q_{3j}+R_{31}+R_{32}+1}}}...\eta^{q_{\sum_{j}Q_{4j}+R_{41}+R_{42}+R_{45}+R_{43}}|t_{\sum_j{Q_{3j}+R_{31}}+R_{32}+R_{34}}}
\cr
\eta^{r_{\sum_{j}Q_{5j}+1+R_{51}+R_{52}+R_{54}}|t_{\sum_j{Q_{3j}+R_{31}+R_{32}+R_{34}+1}}}...\eta^{r_{\sum{s}-3}|t_{s_3-1}}
}}

where in our notations
$\eta^{m_i|n_j}$ is Minkowski tensor and $\sum_{j}A_{ij}\equiv\sum_{j=1;j\neq{i}}^5A_{ij}$

This  concludes the $X$-part evaluation.
Next, consider the ghost part of the amplitude.
The relevant correlator is
\eqn\grav{\eqalign{
A_{gh}(w;z;\xi)=
<ce^{-s_1\phi}(w)ce^{-s_2\phi}(1)e^{(u-2)\phi}B^{(2u-2)}_{2\phi-2\chi-\sigma}(z)
e^{(\Sigma{s}-u)\phi}B^{(2\Sigma{s}-u)}_{2\phi-2\chi-\sigma}(\xi)
ce^{-s_3\phi}(0)>}}
To calculate it, introduce again the characteristic partitions:
\eqn\grav{\eqalign{
2(u-2)=4=\alpha_1+\alpha_2+\alpha_3+\alpha_4^{(1)}+\alpha_4^{(2)}
\cr
2(\Sigma{s}-u){\equiv}2(\Sigma{s}-4)=\beta_1+\beta_2+\beta_3+\beta_4^{(1)}+\beta_4^{(2)}}}
where the integers $\alpha_j(j=1,2,3)$ refer to the OPE singularity orders
$\sim(z-z_j)^{-\alpha_j}$ due to contractions
of $B^{(2u-2)}_{2\phi-2\chi-\sigma}(z)$ with the ghost exponents $ce^{-s_j\phi}(z_j)$;
$\beta_j(j=1,2,3)$ refer to the OPE singularities $\sim(\xi-z_j)^{-\beta_j}$
due to contractions
of $B^{(2\Sigma{s}-u)}_{2\phi-2\chi-\sigma}(\xi)$ with $ce^{-s_j\phi}(z_j)$;
$\alpha_4^{(1)}$ $\beta_4^{(1)}$ refer to the OPE singularities of the Bell polynomials
at $z$ and $\xi$ with the opposite ghost exponents $e^{(\Sigma{s}-u)\phi}(\xi)$
and $B^{(2u-2)}_{2\phi-2\chi-\sigma}(z)$ respectively and, finally
$\alpha_4^{(2)}$ and $\beta_4^{(2)}$ refer to the OPE due to contractions between Bell polynomials
at $z$ and $\xi$, i.e.
\eqn\grav{\eqalign{
B^{(2u-2)}_{2\phi-2\chi-\sigma}(z)B^{(2\Sigma{s}-u)}_{2\phi-2\chi-\sigma}(\xi)
\cr
\sim
\sum_{\alpha_4^{(2)},\beta_4^{(2)}}(z-\xi)^{-\alpha_4^{(2)}-\beta_4^{(2)}}\lambda(\alpha_4^{(2)},\beta_4^{(2)})
:B^{(2u-2-\alpha_4^{(2)})}_{2\phi-2\chi-\sigma}(z)B^{(2\Sigma{s}-u-\beta_4^{(2)})}_{2\phi-2\chi-\sigma}(\xi):
}}
where $\lambda(\alpha_4^{(2)},\beta_4^{(2)})$ are the structure constants that can be computed (see below).
The correlator (23) is thus equal to the universal
 factor due to contractions between the ghost exponents,
multiplied by the sum stemming from contractions of the Bell polynomials with themselves
and with various exponents, with each term in the sum 
corresponding to some of the partitions (24), with the
correlator being given by the sum over the partitions.
The explicit expressions for the operator products, needed to compute the correlator, are
~{\selfsft}
\eqn\grav{\eqalign{
B_{{\vec{\alpha}}{\vec{\varphi}}}^{(N_1)}(z)B_{{\vec{\beta}}{\vec{\varphi}}}^{(N_2)}(w)
=\sum_{n_1=0}^{N_1}\sum_{n_2=0}^{N_2}(z-w)^{-n_1-n_2}\lambda(n_1,n_2)
:B_{{\vec{\alpha}}{\vec{\varphi}}}^{(N_1)}(z)B_{{\vec{\beta}}{\vec{\varphi}}}^{(N_2)}(w):
}}
and
\eqn\grav{\eqalign{
B_{{\vec{\alpha}}{\vec{\varphi}}}^{(N)}(z)e^{{{\vec{\beta}}{\vec{\varphi}}}}(w)
=\sum_{n=0}^{N}(z-w)^{-n}{{\Gamma(-{\vec{\alpha}}{\vec{\beta}}+1)}
\over{n!\Gamma(-{\vec{\alpha}}{\vec{\beta}}+1-n)}}
:B_{{\vec{\alpha}}{\vec{\varphi}}}^{(N-n)}(z)e^{{{\vec{\beta}}{\vec{\varphi}}}}(w):}}
where
\eqn\grav{\eqalign{{{\vec{\alpha}}{\vec{\varphi}}}\equiv
\alpha_1\phi+\alpha_2\chi+\alpha_3\sigma\cr
{{\vec{\alpha}}{\vec{\beta}}}=\alpha_1\beta_1-\alpha_2\beta_2-\alpha_3\beta_3}}
and the OPE structure constants in (26) are given by ~{\selfsft}
\eqn\grav{\eqalign{\lambda(n_1,n_2)={1\over{n_1!n_2!}}\partial_x^{n_1}
\partial_y^{n_2}F_\lambda(x,y)|_{x=y=0}\cr
F_\lambda(x,y)=\lbrack{{(1+x)(1-y)}\over{1+x-y}}\rbrack^{{\vec{\alpha}}{\vec{\beta}}}}}
Clearly, in our case
${{\vec{\alpha}}{\vec{\beta}}}=-1$
so one simply has
\eqn\lowen{\lambda(n_1,n_2)=(-1)^{n_1}}
With all the above identities it is now
straightforward to compute the ghost correlator, with the result given by:
\eqn\grav{\eqalign{
A_{gh}(w;z;\xi)
\cr=
\sum_{{{\lbrack}partitions}:2(u-2){\equiv}4|\alpha_1+\alpha_2+\alpha_3+\alpha_4^{(1)}+\alpha_4^{(2)}\rbrack}
\cr
\sum_{{{\lbrack}partitions}:2(\Sigma{s}-u){\equiv}2(\Sigma{s}-4)|\beta_1+\beta_2+\beta_3+\beta_4^{(1)}+\beta_4^{(2)}\rbrack}
\cr
{\prod_{j=1}^3}{{((2s_j-1)!)^2}\over{\alpha_j!\beta_j!(2s_j-1-\alpha_j)!(2s_j-1-\beta_j)!}}
\cr\times
{{\Gamma(3-2u)\Gamma(1+2u-2\Sigma{s})}\over{\Gamma(3-2u-\alpha_4^{(1)})
\Gamma(1+2u-2\Sigma{s}-\beta_4^{(1)})}}|_{u\rightarrow{4}}
\cr\times
(-1)^{\alpha_4^{(2)}+1}w^{s_1^2+2-\alpha_1-\beta_1}
(1-z)^{2s_2-\alpha_2}(1-\xi)^{s_2(\Sigma{s}-4)-\beta_2}
\cr\times
z^{2s_3-\alpha_3}\xi^{s_3(\Sigma{s}-4)-\beta_3}
(z-\xi)^{2(4-\Sigma{s})-\alpha_4^{(1)}-\alpha_4^{(2)}-\beta_4^{(1)}-\beta_4^{(2)}}}}
This concludes the computation of the ghost factor, contributing to the integrand
of the 5-point amplitude.
Finally, we are left with computing the $\psi$-factor,
given by
\eqn\grav{\eqalign{
A_\psi(w;z;\xi)=
\cr
<\prod_{j_1=0}^{s_1-3}\partial^{j_1}\psi_{\alpha_{j_1}}(w)
\prod_{j_2=0}^{s_2-3}\partial^{j_2}\psi_{\beta_{j_2}}(1)
\prod_{j_3=0}^{u-3}\partial^{j_3}\psi_{\gamma_{j_3}}(z)
\prod_{j_4=0}^{\Sigma{s}-u-1}\partial^{j_4}\psi_{\lambda_{j_4}}(\xi)
\prod_{j_5=0}^{s_3-3}\partial^{j_5}\psi_{\sigma_{j_5}}(0)>}}
This amplitude would be actually the most tedious one to compute 
for generic spin values, since there seems to be no way to systemize 
it in terms of sums over partitions of the spin values, as it has been done
for the previous correlators.
However, assuming that $u$ is the minimal spin value of the operators,
 it simplifies significantly for the minimal possible $u=4$ value (for $u<4$
it vanishes identically).
The simplification is that for $u=4$, all the $\psi$-fields of 4 operators
have to couple to the $\psi$-fields of the operator with the highest spin value,
given by $s_{max}=\Sigma{s}-2$.  Technically, this means that
in the corresponding space-time amplitude all the 
4 out of 5 $\Omega^{s-1|t}$
extra-field's $t$-indices $\alpha,\beta,\gamma,\sigma$ 
have to contract to the $\lambda$-index of the highest spin field
$\Omega^{s_{max}-1|\lambda}$.
The computation performed below is still possible to perform for $u>4$, 
e.g. for $u=5$ or 6, however, for $u>4$ the locality of the interaction vertex in space-time
would be broken by the RG flows from lower orders despite the locality of the scattering amplitude
(see the discussion in the Section 4)
In case of $u=4$, 
it is convenient to make the following definition.
Let $i(\alpha_k)$ describe the  contraction of the worldsheet fermion
$\partial^{k}\psi_{\alpha_k}$ of the spin $s_1$ operator to the fermion
$\partial^{i(\alpha_k)}\psi_{\lambda_{i(\alpha_k)}}$ of the spin $s_{max}$ operator.
Similarly, let $i(\beta_k)$ describe the contractions between the worldsheet fermions
of the $s_2$ and $s_{max}$ operators, and so on.
Clearly, $i(\alpha)\neq{i(\beta)}\neq{i(\gamma)}\neq{i(\sigma)}$
for all values of $\alpha,\beta,\gamma,\sigma$
and $0\leq{i(\alpha,\beta,\gamma,\sigma)}\leq{s_{max}}-3$
 Then it is natural to express the
$\psi$-correlator in terms of the sum over the permutations of different $i$'s , or, equivalently,
in terms of sum over all possible  length $s_{max}-2$
orderings of unequal integer numbers from $0$ to $s_{max}-3$.

The $A_\psi$ correlator is then straightforward to compute in terms of such a sum, with 
the result given by
\eqn\grav{\eqalign{
A_\psi(w,z,\xi)
\cr=
(-1)^{s_1s_2+s_1+s_2+s_3}
\sum_{{\lbrack}permutations:{\lbrace},i(\alpha_0)...i(\alpha_{s_1-3}),i(\beta_0)
...i(\beta_{s_2-3}),i(\gamma_0),i(\gamma_1),i(\sigma_0)...i(\sigma_{s_3-3})\rbrace\rbrack}
\cr\lbrace
(-1)^{\pi({\lbrace},i(\alpha_0)...i(\alpha_{s_1-3}),i(\beta_0)
...i(\beta_{s_2-3}),i(\gamma_0),i(\gamma_1),i(\sigma_0)...i(\sigma_{s_3-3})\rbrace)}
\cr\times
(1+i(\gamma_0))!(2+i(\gamma_1))!
\cr
\prod_{k_1=1}^{s_1-2}(k_1+i(\alpha_{k_1}-1))!
\prod_{k_2=1}^{s_2-2}(k_2+i(\beta_{k_2}-1))!
\cr\times
\prod_{k_3=1}^{s_3-2}(k_3+i(\beta_{k_3}-1))!
w^{{1\over2}(s_1-2)^2+\sum_{l_1=0}^{s_1-3}i(\alpha_{l_1})}
\cr
(1-\xi)^{{1\over2}(s_2-2)^2+\sum_{l_2=0}^{s_2-3}i(\beta_{l_2})}
\xi^{{1\over2}(s_3-2)^2+\sum_{l_3=0}^{s_2-3}i(\sigma_{l_3})}
(z-\xi)^{2+i(\gamma_0)+i(\gamma_1)}
\rbrace}}
where

$${\pi({\lbrace}i(\alpha_0)...i(\alpha_{s_1-3}),i(\beta_0)
...i(\beta_{s_2-3}),i(\gamma_0),i(\gamma_1),i(\sigma_0)...i(\sigma_{s_3-3})\rbrace)}$$

is the minimal number of the permutations it takes to convert
the ordering

$${\lbrace}i(\alpha_0)...i(\alpha_{s_1-3}),i(\beta_0)...i(\beta_{s_2-3}),i(\gamma_0),i(\gamma_1),i(\sigma_0)...i(\sigma_{s_3-3})\rbrace$$

in the argument of $\pi$ into the reference ordering
$$\lbrace{0,1,2,.....,{s_{max}-3}}\rbrace$$  of $s_{max}-2$ integers.
This concludes the computation of the $\psi$-factor contribuiting to the 
integrand of the amplitude. The final remaining step
to determine the amplitude describing the $5$-point higher spin interaction
is to perform the double worldsheet integration
over the positions of the integrated vertices, given by
\eqn\grav{\eqalign{A(p_1,...,p_5)=
\cr
\int_0^1{dz}\int_0^z{d\xi}z^{4}\xi^{2(\Sigma{s}-4)}
A_\psi(w;z;\xi)
A_{gh}(w;z;\xi)
A_X(p_1,...p_5|w;z;\xi)}}
to set the limit $w\rightarrow{\infty}$ and to 
contract the result with the space-time frame-like higher spin fields
(in the integral (9) it is convenient to choose the reference $u$-points
$u=0$ in the  the both of the homotopy transformations
for the integrated vertex operators ).
Combining $A_{X},A_\psi$ and $A_{gh}$ factors together, evaluating the integral
(34) and contracting with the frame-like space-time fields, we obtain the
overall quintic amplitude, given by
\eqn\grav{\eqalign{
A(p_1,...,p_5)=6\pi\Omega_{m_1...m_{s_1-1}|\lambda_0...\lambda_{s_1-3}}(p_5)
\Omega_{n_1...n_{s_2-1}|\lambda_{s_1-2}...\lambda_{s_1+s_2-5}}(p_4)
\cr
\Omega_{q_1q_2q_3|\lambda_{s_1+s_2-4}\lambda_{s_1+s_2-3}}(p_3)
\Omega_{r_1...r_{\Sigma{s}-3}}^{\lambda_0...\lambda_{\Sigma{s}-5}}(p_2)
\Omega_{t_1...t_{s_3-1}|\lambda_{s_1+s_2-2}...\lambda_{\Sigma{s}-5}}(p_1)
\cr\times
\sum_{{partitions:}s_i-1|\sum_{j=1,j\neq{i}}^5(R_{ij}+Q_{ij});i=1,...,5}
\cr
\sum_{{partitions}:2(u-2){\equiv}4|\alpha_1+\alpha_2+\alpha_3+\alpha_4^{(1)}+\alpha_4^{(2)}}
\cr
\sum_{{partitions}:2(\Sigma{s}-u){\equiv}2(\Sigma{s}-4)|\beta_1+\beta_2+\beta_3+\beta_4^{(1)}+\beta_4^{(2)}}
\cr
\sum_{permutations:{\lbrace},i(\alpha_0)...i(\alpha_{s_1-3}),i(\beta_0)
...i(\beta_{s_2-3}),i(\gamma_0),i(\gamma_1),i(\sigma_0)...i(\sigma_{s_3-3})\rbrace}
\cr\lbrace
(-1)^{\pi({\lbrace},i(\alpha_0)...i(\alpha_{s_1-3}),i(\beta_0)
...i(\beta_{s_2-3}),i(\gamma_0),i(\gamma_1),i(\sigma_0)...i(\sigma_{s_3-3})\rbrace)+s_1s_2+s_1+s_2+s_3+\alpha_4^{(2)}}
\cr
(1+i(\gamma_0))!(2+i(\gamma_1))!
\prod_{k_1=1}^{s_1-2}(k_1+i(\alpha_{k_1}-1))!
\cr
\prod_{k_2=1}^{s_2-2}(k_2+i(\beta_{k_2}-1))!
\prod_{k_3=1}^{s_3-2}(k_3+i(\beta_{k_3}-1))!
\cr\times
{\prod_{j=1}^3}{{((2s_j-1)!)^2}\over{\alpha_j!\beta_j!(2s_j-1-\alpha_j)!(2s_j-1-\beta_j)!}}
\cr\times
{{\Gamma(3-2u)\Gamma(1+2u-2\Sigma{s})}\over{\Gamma(3-2u-\alpha_4^{(1)})
\Gamma(1+2u-2\Sigma{s}-\beta_4^{(1)})}}|_{u\rightarrow{4}}
\cr\times
{{(s_1-1)!(s_2-1)!(s_3-1)!(\sum{s}-3)!}\over{
{\prod_{i=1}^4\prod_{j=2;j>i}^5R_{ij}!\prod_{k=1}^5\prod_{l=1;k{\neq}l}^5
Q_{kl}!}}} 
\cr
(i)^{\sum_{i,j=1;i\neq{j}}^5Q_{ij}}
(-1)^{\sum_{j=1}^5(Q_{1j}+Q_{2j})-Q_{21}+Q_{43}+Q_{43}+Q_{45}+Q_{53}}
\cr\times
p_1^{m_{Q_{12}+Q_{14}+Q_{15}+1}}...p_1^{m_{Q_{12}+Q_{14}+Q_{15}+Q_{13}}}
p_1^{n_{Q_{21}+Q_{24}+Q_{25}+1}}...p_1^{n_{Q_{21}+Q_{24}+Q_{25}+Q_{23}}}
\cr
p_1^{q_{Q_{41}+Q_{42}+Q_{45}+1}}...p_1^{q_{Q_{41}+Q_{42}+Q_{45}+Q_{43}}}
p_1^{r_{Q_{51}+Q_{52}+Q_{54}+1}}...p_1^{r_{Q_{51}+Q_{52}+Q_{54}+Q_{53}}}
\cr
p_2^{m_{Q_{12}+Q_{14}+1}}...p_2^{m_{Q_{12}+Q_{14}+Q_{15}}}
p_2^{n_{Q_{21}+Q_{24}+1}}...p_2^{n_{Q_{21}+Q_{24}+Q_{25}}}
\cr
p_2^{q_{Q_{41}+Q_{42}+1}}...p_2^{q_{Q_{41}+Q_{42}+Q_{45}}}
p_2^{t_{Q_{31}+Q_{32}+Q_{34}+1}}...p_2^{t_{Q_{31}+Q_{32}+Q_{34}+Q_{35}}}
\cr
p_3^{m_{Q_{12}+1}}...p_3^{m_{Q_{12}+Q_{14}}}
p_3^{n_{Q_{21}+1}}...p_3^{n_{Q_{21}+Q_{24}}}
\cr
p_3^{r_{Q_{51}+Q_{52}+1}}...p_3^{r_{Q_{51}+Q_{52}+Q_{54}}}
p_3^{t_{Q_{31}+Q_{32}+1}}...p_3^{t_{Q_{31}+Q_{32}+Q_{34}}}
}}
\vfill\eject
\eqn\grav{\eqalign{{\times}
p_4^{m_{1}}...p_4^{m_{Q_{12}}}
p_4^{q_{Q_{41}+1}}...p_4^{q_{Q_{41}+Q_{42}}}
p_4^{r_{Q_{51}+1}}...p_4^{r_{Q_{51}+Q_{52}}}
p_4^{t_{Q_{31}+1}}...p_4^{t_{Q_{31}+Q_{32}}}
\cr
p_5^{n_{1}}...p_5^{n_{Q_{21}}}
p_5^{q_{1}}...p_5^{q_{Q_{41}}}
p_5^{r_{1}}...p_5^{r_{Q_{51}}}
p_5^{t_{1}}...p_5^{t_{Q_{31}}}
\cr\times
{{\Gamma{\lbrack}p_2p_3+9-T_{45}-2\Sigma{s}-\beta_2-{1\over2}(s_2-2)^2-\sum_{k=1}^{s_2-2}i(\beta_{k-1})\rbrack}
\over{\Gamma{\lbrack}-p_2p_4+T_{25}-s_2(\Sigma{s}-4)+{1\over2}(s_2-2)^2+\beta_2+\sum_{k=1}^{s_2-2}i(\beta_{k-1})\rbrack}}
\cr\times
{{\Gamma{\lbrack}p_3p_4-T_{24}+2s_2-\alpha_2\rbrack}\over{sin{\lbrack}\pi(p_2p_4-T_{25}+s_2(\Sigma{s}-4)
-{1\over2}(s_2-2)^2-\beta_2-\sum_{k=1}^{s_2-2}i(\beta_{k-1})){\rbrack}}}
\cr\times
\Gamma{\lbrack}p_2p_4+p_2p_3+p_1p_3+6-T_{25}-T_{45}-T_{34}+
\cr
(s_2-2)(\Sigma{s}-4)-{1\over2}(s_2-2)^2+2s_3-\alpha_3
-\sum_{k=1}^{s_2-2}i(\beta_{k-1})-i(\gamma_0)-i(\gamma_1)\rbrack
\cr
\Gamma^{-1}{\lbrack}p_2p_4+p_2p_3+2-T_{25}-T_{45}+(s_2-2)(\Sigma{s}-4)
\cr
-{1\over2}(s_2-2)^2
-\sum_{k=1}^{s_2-2}i(\beta_{k-1})-i(\gamma_0)-i(\gamma_1){\rbrack}
\cr\times
\Gamma{\lbrack}p_2p_4+p_2p_3+p_1p_3+p_3p_4+6-T_{24}+2s_2-\alpha_2-T_{25}-T_{45}-T_{34}
\cr
+(s_2-2)(\Sigma{s}-4)-{1\over2}(s_2-2)^2+2s_3-\alpha_3
-\sum_{k=1}^{s_2-2}i(\beta_{k-1})-i(\gamma_0)-i(\gamma_1)\rbrack
\cr\times
_3F_2{\lbrack}-p_1p_2+T_{35}-(s_3+2)(\Sigma{s}-4)+{1\over2}(s_3-2)^2+\beta_3+\sum_{k=1}^{s_3-2}i(\beta_{k-1});
\cr
p_2p_4+1-T_{25}+s_2(\Sigma{s}-4)
-{1\over2}(s_2-2)^2-\beta_2-\sum_{k=1}^{s_2-2}i(\beta_{k-1});
\cr
p_2p_4+p_2p_3+p_1p_3+6-T_{25}-T_{45}-T_{34}
+(s_2-2)(\Sigma{s}-4)
\cr
-{1\over2}(s_2-2)^2+2s_3-\alpha_3
-\sum_{k=1}^{s_2-2}i(\beta_{k-1})-i(\gamma_0)-i(\gamma_1);
\cr 
p_2p_4+p_2p_3+2
-T_{25}-T_{45}+(s_2-2)(\Sigma{s}-4)
\cr
-{1\over2}(s_2-2)^2
-\sum_{k=1}^{s_2-2}i(\beta_{k-1})-i(\gamma_0)-i(\gamma_1);
\cr
p_2p_4+p_2p_3+p_1p_3+p_3p_4+6
-T_{25}-T_{45}-T_{34}-T_{24}
+(s_2-2)(\Sigma{s}-4)
\cr
-{1\over2}(s_2-2)^2+2s_2+2s_3
-\alpha_2-\alpha_3
-\sum_{k=1}^{s_2-2}i(\beta_{k-1})-i(\gamma_0)-i(\gamma_1); 1\rbrack
}}
\vfill\eject
\eqn\grav{\eqalign{
\times
\eta^{m_{\sum_{j}Q_{1j}+1}|n_{\sum_j{Q_{2j}+1}}}...\eta^{m_{\sum_{j}Q_{1j}+R_{12}}|n_{\sum_j{Q_{2j}+R_{12}}}}
\cr
\eta^{m_{\sum_{j}Q_{1j}+1+R_{12}}|q_{\sum_j{Q_{4j}+1}}}...\eta^{m_{\sum_{j}Q_{1j}+R_{12}+R_{14}}|q_{\sum_j{Q_{4j}+R_{14}}}}
\cr
\eta^{m_{\sum_{j}Q_{1j}+R_{12}+R_{14}+1}|r_{\sum_j{Q_{5j}+1}}}...\eta^{m_{\sum_{j}Q_{1j}+R_{12}+R_{14}+R_{15}}|r_{\sum_j{Q_{5j}+R_{15}}}}
\cr
\eta^{m_{\sum_{j}Q_{1j}+R_{12}+R_{14}+R_{15}+1}|t_{\sum_j{Q_{3j}+1}}}...\eta^{m_{\sum_{j}Q_{1j}+R_{12}+R_{14}+R_{15}+R_{13}}|t_{\sum_j{Q_{3j}+R_{13}}}}
\cr
\eta^{n_{\sum_{j}Q_{2j}+1+R_{12}}|q_{\sum_j{Q_{4j}+R_{14}+1}}}...\eta^{m_{\sum_{j}Q_{2j}+R_{12}+R_{24}}|q_{\sum_j{Q_{4j}+R_{14}}+R_{24}}}
\cr
\eta^{n_{\sum_{j}Q_{2j}+1+R_{12}+R_{24}}|r_{\sum_j{Q_{5j}+R_{15}+1}}}...\eta^{m_{\sum_{j}Q_{2j}+R_{12}+R_{24}+R_{25}}|r_{\sum_j{Q_{5j}+R_{15}}+R_{25}}}
\cr
\eta^{n_{\sum_{j}Q_{2j}+1+R_{12}+R_{24}+R_{25}+1}|t_{\sum_j{Q_{3j}+R_{31}+1}}}...\eta^{m_{\sum_{j}Q_{2j}+R_{12}+R_{24}+R_{25}}|t_{\sum_j{Q_{3j}+R_{31}}+R_{32}}}
\cr
\eta^{q_{\sum_{j}Q_{4j}+1+R_{41}+R_{42}}|r_{\sum_j{Q_{5j}+R_{51}+R_{52}+1}}}...\eta^{q_{\sum_{j}Q_{4j}+R_{41}+R_{42}+R_{45}}|r_{\sum_j{Q_{5j}+R_{15}}+R_{25}+R_{54}}}
\cr
\eta^{q_{\sum_{j}Q_{4j}+1+R_{41}+R_{42}+R_{45}}|t_{\sum_j{Q_{3j}+R_{31}+R_{32}+1}}}...\eta^{q_{\sum_{j}Q_{4j}+R_{41}+R_{42}+R_{45}+R_{43}}|t_{\sum_j{Q_{3j}+R_{31}}+R_{32}+R_{34}}}\cr
\eta^{r_{\sum_{j}Q_{5j}+1+R_{51}+R_{52}+R_{54}}|t_{\sum_j{Q_{3j}+R_{31}+R_{32}+R_{34}+1}}}...\eta^{r_{\sum{s}-3}|t_{s_3-1}}
\rbrace}}
where 
\eqn\lowen{T_{ij}=2R_{ij}+Q_{ij}+Q_{ji}}

and the additional constraint is imposed on the partitions:

\eqn\grav{\sum_{j=2}^5Q_{j1}+\sum_{k=1}^{s_1-2}i(\alpha_{k-1})=3+{{s_1(s_1-1)}\over2}}

stemming from the fact that only $w^{0}$-terms conmtribute to the amplitude, with all others vanishing
in the limit $w\rightarrow{\infty}$.
This concludes the computation of the five-point amplitude for the localized quintic interaction.
In the next section we shall discuss the construction of the quintic higher spin vertex, related to this amplitude.

\centerline{\bf 4. Reading off the Quintic Vertex}

In the previous section we have computed the five-point worldsheet amplitude, related to the 
interaction of masssless higher spin in the quintic order. By itself, this amplitude does not 
describe yet the 5-point interaction vertex in the low energy effective action: it is only gauge-invariant
under gauge (BRST) transformations at the linearized level. To read off the 
interaction vertex with the full gauge symmetry, one  has to subtract from it the terms , 
produced as a result of the worldsheet RG flows of the effective action's terms at lower orders, such as 
cubic and quartic. Generally speaking, there are two possible sources of such terms at the quintic order:
the flow of the quartic vertex in the leading $\alpha^\prime$ order and the flow of the cubic vertex
in the subleading $\alpha^\prime$ order. In general, computation of these flow terms would be quite complicated;
in particular they would involve the flows stemming from all the diversity of the quartic vertices which by
themselves are tedious and complex. Moreover, since the quartic interactions are generally nonlocal, one
would generally expect the flow terms to retain these nonlocalities, 
destroying the local structure
of the amplitude  (35)-(37). 
Fortunately, however, for the spin combinations considered in our work
things again get drastically
simplified: as we pointed out above, the four-point correlation functions
contributing to the worldsheet $\beta$-function of the space-time frame-like field
with the highest spin value , 
relevant to the flow terms at the quintic order, vanish identically,
as it is impossible to accommodate 
the full contractions of the $\psi$-fermions consistently with the
ghost number balance restrictions. 
Therefore the quartic interactions, relevant to the flow terms
at the quintic order, stem themselves from the flows from the previous (cubic) order.
For this reason, the   flow contributions are reduced to the double composition of the RG flows
of the cubic terms which structure is relatively simple to control. As the structure of the 
cubic higher spin vertices is determined by the structure constants of  the higher spin algebra
~{\selfframe, \selfframes}
 and the quintic vertex is cubic in the structure constants,
 the effect of the cubic terms flows
can be expressed by regularizing the poles in $s_{ij}$-channels
present in the amplitude (35)-(37) due to the Euler's 
gamma-functions and the hypergeometric function
where 
\eqn\grav{\eqalign{
s_{12}={1\over2}(p_1-p_2)^2;
s_{13}={1\over2}(p_1-p_3)^2;
s_{14}={1\over2}(p_1-p_4)^2\cr
s_{23}={1\over2}(p_2-p_3)^2;
s_{24}={1\over2}(p_2-p_4)^2;
s_{34}={1\over2}(p_3+p_4)^2\cr
\sum_{i,j}s_{ij}=0}}
are the generalized Mandelstam variables.
Subtracting the poles resulting from the flows of the cubic vertices using the procedure 
similar to those explained in ~{\tseytlin, \selfquartic}
 and taking the field theory limit we find the 
quintic vertex  stemming from the $\beta$-function of the highest spin field is given by:
\eqn\grav{\eqalign{
A(p_1,...,p_5)=6\pi\Omega_{m_1...m_{s_1-1}|\lambda_0...\lambda_{s_1-3}}(p_5)
\Omega_{n_1...n_{s_2-1}|\lambda_{s_1-2}...\lambda_{s_1+s_2-5}}(p_4)
\cr
\Omega_{q_1q_2q_3|\lambda_{s_1+s_2-4}\lambda_{s_1+s_2-3}}(p_3)
\Omega_{r_1...r_{\Sigma{s}-3}}^{\lambda_0...\lambda_{\Sigma{s}-5}}(p_2)
\Omega_{t_1...t_{s_3-1}|\lambda_{s_1+s_2-2}...\lambda_{\Sigma{s}-5}}(p_1)
\cr\times
\sum_{{partitions:}s_i-1|\sum_{j=1,j\neq{i}}^5(R_{ij}+Q_{ij});i=1,...,5}
\cr
\sum_{{partitions}:2(u-2){\equiv}4|\alpha_1+\alpha_2+\alpha_3+\alpha_4^{(1)}+\alpha_4^{(2)}}
\cr
\sum_{{partitions}:2(\Sigma{s}-u){\equiv}2(\Sigma{s}-4)|\beta_1+\beta_2+\beta_3+\beta_4^{(1)}+\beta_4^{(2)}}
\cr
\sum_{permutations:{\lbrace},i(\alpha_0)...i(\alpha_{s_1-3}),i(\beta_0)
...i(\beta_{s_2-3}),i(\gamma_0),i(\gamma_1),i(\sigma_0)...i(\sigma_{s_3-3})\rbrace}
\cr\lbrace
(-1)^{\pi({\lbrace},i(\alpha_0)...i(\alpha_{s_1-3}),i(\beta_0)
...i(\beta_{s_2-3}),i(\gamma_0),i(\gamma_1),i(\sigma_0)...i(\sigma_{s_3-3})\rbrace)+s_1s_2+s_1+s_2+s_3+\alpha_4^{(2)}}
\cr
(1+i(\gamma_0))!(2+i(\gamma_1))!
\prod_{k_1=1}^{s_1-2}(k_1+i(\alpha_{k_1}-1))!
\cr
\prod_{k_2=1}^{s_2-2}(k_2+i(\beta_{k_2}-1))!
\prod_{k_3=1}^{s_3-2}(k_3+i(\beta_{k_3}-1))!
\cr\times
{\prod_{j=1}^3}{{((2s_j-1)!)^2}\over{\alpha_j!\beta_j!(2s_j-1-\alpha_j)!(2s_j-1-\beta_j)!}}
\cr\times
{{\Gamma(3-2u)\Gamma(1+2u-2\Sigma{s})}\over{\Gamma(3-2u-\alpha_4^{(1)})
\Gamma(1+2u-2\Sigma{s}-\beta_4^{(1)})}}|_{u\rightarrow{4}}
\cr\times
{{(s_1-1)!(s_2-1)!(s_3-1)!(\sum{s}-3)!}\over{
{\prod_{i=1}^4\prod_{j=2;j>i}^5R_{ij}!\prod_{k=1}^5\prod_{l=1;k{\neq}l}^5
Q_{kl}!}}} 
\cr
(i)^{\sum_{i,j=1;i\neq{j}}^5Q_{ij}}
(-1)^{\sum_{j=1}^5(Q_{1j}+Q_{2j})-Q_{21}+Q_{43}+Q_{43}+Q_{45}+Q_{53}}
\cr\times
p_1^{m_{Q_{12}+Q_{14}+Q_{15}+1}}...p_1^{m_{Q_{12}+Q_{14}+Q_{15}+Q_{13}}}
p_1^{n_{Q_{21}+Q_{24}+Q_{25}+1}}...p_1^{n_{Q_{21}+Q_{24}+Q_{25}+Q_{23}}}
\cr
p_1^{q_{Q_{41}+Q_{42}+Q_{45}+1}}...p_1^{q_{Q_{41}+Q_{42}+Q_{45}+Q_{43}}}
p_1^{r_{Q_{51}+Q_{52}+Q_{54}+1}}...p_1^{r_{Q_{51}+Q_{52}+Q_{54}+Q_{53}}}
\cr
p_2^{m_{Q_{12}+Q_{14}+1}}...p_2^{m_{Q_{12}+Q_{14}+Q_{15}}}
p_2^{n_{Q_{21}+Q_{24}+1}}...p_2^{n_{Q_{21}+Q_{24}+Q_{25}}}
\cr
p_2^{q_{Q_{41}+Q_{42}+1}}...p_2^{q_{Q_{41}+Q_{42}+Q_{45}}}
p_2^{t_{Q_{31}+Q_{32}+Q_{34}+1}}...p_2^{t_{Q_{31}+Q_{32}+Q_{34}+Q_{35}}}
\cr
p_3^{m_{Q_{12}+1}}...p_3^{m_{Q_{12}+Q_{14}}}
p_3^{n_{Q_{21}+1}}...p_3^{n_{Q_{21}+Q_{24}}}
\cr
p_3^{r_{Q_{51}+Q_{52}+1}}...p_3^{r_{Q_{51}+Q_{52}+Q_{54}}}
p_3^{t_{Q_{31}+Q_{32}+1}}...p_3^{t_{Q_{31}+Q_{32}+Q_{34}}}
}}
\vfill\eject
\eqn\grav{\eqalign{{\times}
p_4^{m_{1}}...p_4^{m_{Q_{12}}}
p_4^{q_{Q_{41}+1}}...p_4^{q_{Q_{41}+Q_{42}}}
p_4^{r_{Q_{51}+1}}...p_4^{r_{Q_{51}+Q_{52}}}
p_4^{t_{Q_{31}+1}}...p_4^{t_{Q_{31}+Q_{32}}}
\cr
p_5^{n_{1}}...p_5^{n_{Q_{21}}}
p_5^{q_{1}}...p_5^{q_{Q_{41}}}
p_5^{r_{1}}...p_5^{r_{Q_{51}}}
p_5^{t_{1}}...p_5^{t_{Q_{31}}}
\cr\times
{\lbrack}(-9+T_{45}+2\Sigma{s}+\beta_2+{1\over2}(s_2-2)^2+\sum_{k=1}^{s_2-2}i(\beta_{k-1}))!\rbrack^{-1}
\cr\times
L(-9+T_{45}+2\Sigma{s}+\beta_2+{1\over2}(s_2-2)^2+\sum_{k=1}^{s_2-2}i(\beta_{k-1}))
\cr\times
(-T_{25}+s_2(\Sigma{s}-4)
-{1\over2}(s_2-2)^2-\beta_2-\sum_{k=1}^{s_2-2}i(\beta_{k-1}))!
(1-T_{24}+2s_2-\alpha_2)!
\cr\times
{\lbrack}5-T_{25}-T_{45}-T_{34}+
(s_2-2)(\Sigma{s}-4)-{1\over2}(s_2-2)^2
\cr
+2s_3-\alpha_3
-\sum_{k=1}^{s_2-2}i(\beta_{k-1})-i(\gamma_0)-i(\gamma_1){\rbrack}!
\cr
\lbrack
(1-T_{25}-T_{45}+(s_2-2)(\Sigma{s}-4)
-{1\over2}(s_2-2)^2
-\sum_{k=1}^{s_2-2}i(\beta_{k-1})-i(\gamma_0)-i(\gamma_1))!\rbrack^{-1}
\cr\times
{\lbrack}5-T_{24}+2s_2-\alpha_2-T_{25}-T_{45}-T_{34}
+(s_2-2)(\Sigma{s}-4)-{1\over2}(s_2-2)^2
\cr
+2s_3-\alpha_3
-\sum_{k=1}^{s_2-2}i(\beta_{k-1})-i(\gamma_0)-i(\gamma_1){\rbrack}!
\cr\times
_3F_2{\lbrack}T_{35}-(s_3+2)(\Sigma{s}-4)+{1\over2}(s_3-2)^2+\beta_3+\sum_{k=1}^{s_3-2}i(\beta_{k-1});
\cr
1-T_{25}+s_2(\Sigma{s}-4)
-{1\over2}(s_2-2)^2-\beta_2-\sum_{k=1}^{s_2-2}i(\beta_{k-1});
\cr
6-T_{25}-T_{45}-T_{34}
+(s_2-2)(\Sigma{s}-4)
-{1\over2}(s_2-2)^2
\cr
+2s_3-\alpha_3
-\sum_{k=1}^{s_2-2}i(\beta_{k-1})-i(\gamma_0)-i(\gamma_1);
\cr 
2-T_{25}-T_{45}+(s_2-2)(\Sigma{s}-4)
-{1\over2}(s_2-2)^2
-\sum_{k=1}^{s_2-2}i(\beta_{k-1})-i(\gamma_0)-i(\gamma_1);
\cr
6-T_{25}-T_{45}-T_{34}-T_{24}
+(s_2-2)(\Sigma{s}-4)
-{1\over2}(s_2-2)^2+2s_2+2s_3
\cr
-\alpha_2-\alpha_3
-\sum_{k=1}^{s_2-2}i(\beta_{k-1})-i(\gamma_0)-i(\gamma_1); 1\rbrack
}}
\vfill\eject
\eqn\grav{\eqalign{
\times
\eta^{m_{\sum_{j}Q_{1j}+1}|n_{\sum_j{Q_{2j}+1}}}...\eta^{m_{\sum_{j}Q_{1j}+R_{12}}|n_{\sum_j{Q_{2j}+R_{12}}}}
\cr
\eta^{m_{\sum_{j}Q_{1j}+1+R_{12}}|q_{\sum_j{Q_{4j}+1}}}...\eta^{m_{\sum_{j}Q_{1j}+R_{12}+R_{14}}|q_{\sum_j{Q_{4j}+R_{14}}}}
\cr
\eta^{m_{\sum_{j}Q_{1j}+R_{12}+R_{14}+1}|r_{\sum_j{Q_{5j}+1}}}...\eta^{m_{\sum_{j}Q_{1j}+R_{12}+R_{14}+R_{15}}|r_{\sum_j{Q_{5j}+R_{15}}}}
\cr
\eta^{m_{\sum_{j}Q_{1j}+R_{12}+R_{14}+R_{15}+1}|t_{\sum_j{Q_{3j}+1}}}...\eta^{m_{\sum_{j}Q_{1j}+R_{12}+R_{14}+R_{15}+R_{13}}|t_{\sum_j{Q_{3j}+R_{13}}}}
\cr
\eta^{n_{\sum_{j}Q_{2j}+1+R_{12}}|q_{\sum_j{Q_{4j}+R_{14}+1}}}...\eta^{m_{\sum_{j}Q_{2j}+R_{12}+R_{24}}|q_{\sum_j{Q_{4j}+R_{14}}+R_{24}}}
\cr
\eta^{n_{\sum_{j}Q_{2j}+1+R_{12}+R_{24}}|r_{\sum_j{Q_{5j}+R_{15}+1}}}...\eta^{m_{\sum_{j}Q_{2j}+R_{12}+R_{24}+R_{25}}|r_{\sum_j{Q_{5j}+R_{15}}+R_{25}}}
\cr
\eta^{n_{\sum_{j}Q_{2j}+1+R_{12}+R_{24}+R_{25}+1}|t_{\sum_j{Q_{3j}+R_{31}+1}}}...\eta^{m_{\sum_{j}Q_{2j}+R_{12}+R_{24}+R_{25}}|t_{\sum_j{Q_{3j}+R_{31}}+R_{32}}}
\cr
\eta^{q_{\sum_{j}Q_{4j}+1+R_{41}+R_{42}}|r_{\sum_j{Q_{5j}+R_{51}+R_{52}+1}}}...\eta^{q_{\sum_{j}Q_{4j}+R_{41}+R_{42}+R_{45}}|r_{\sum_j{Q_{5j}+R_{15}}+R_{25}+R_{54}}}
\cr
\eta^{q_{\sum_{j}Q_{4j}+1+R_{41}+R_{42}+R_{45}}|t_{\sum_j{Q_{3j}+R_{31}+R_{32}+1}}}...\eta^{q_{\sum_{j}Q_{4j}+R_{41}+R_{42}+R_{45}+R_{43}}|t_{\sum_j{Q_{3j}+R_{31}}+R_{32}+R_{34}}}\cr
\eta^{r_{\sum_{j}Q_{5j}+1+R_{51}+R_{52}+R_{54}}|t_{\sum_j{Q_{3j}+R_{31}+R_{32}+R_{34}+1}}}...\eta^{r_{\sum{s}-3}|t_{s_3-1}}
\rbrace}}
where
\eqn\lowen{L(n)=\sum_{m=1}^n{1\over{m}}}
This concludes the evaluation of the quintic interaction vertex.

\centerline{\bf 5. Conclusion}

In this work we have considered a very special limit of higher spin quintic interaction,
limited to the case when the sum of three spins participating in the interaction roughly equals
the sum of the remaining two, or which one must be small anough and another must be large.
In this case, the ghost structure of the operators drastically simplifies the calculations
and the absence of RG flows from the quartic order, contributing at the fifth order,
makes the construction of the interaction 5-vertex from the scattering amplitude a relatively
straightforward procedure.
Despite the specific choice of the spin values considered in this work, it
is remarkable that the structure of the $5$-vertex can be extracted from string theory
in the low energy limit. One particularly important result is the locality of the quintic amplitude
in this limit. This is the intriguing novelty of the quintic interaction  all the known examples
of the vertices at the previous (quartic) order are essentially nonlocal, and
the vertex (41)-(43) constructed in this paper has no quartic analogue (which vanishes by the ghost/$\psi$-number
constraints). This may bear important important implications for the higher spin holography,
as the nonlocality of higher spin vertices versus 
their local counterparts in the dual CFT's is
the well-known puzzle {\antalf, \antals, \xavierf, \xaviers, \metsaevf}. 
Despite the vanishing of the four-point analogue of the amplitude considered
in this work, one may still hope to find alternative mechanisms for 
localization at the quartic level,
at least in the AdS space.
Our hope is that the string theory may provide efficient and powerful tools
 to address these issues,
as well as to approach the higher orders of higher spin interactions, at least for
specific spin values. It is also of interest to find the holographic interpretation of 
the localization effect, pointed out in our calculations.
For that, it is on the other hand necessary 
to have better understanding of the higher-spin nonlocalities
from the string theory side. So far, string theory 
has been able to account for very limited and simplistic types
of higher spin nonlocalities only. For that, the class of $V_{s-1|s-3}$-operators 
considered in this work, may not be sufficient 
and one may need to find explicit solutions of the
operator equations (15)-(16) for the generalized  zero torsion constraints.
These constraints are generally hard to work with in the formalism  of the on-shell 
(first-quantized) string theory.
Alternatively, in the string field theory (SFT) 
approach the nonlocalities may be naturally encripted
in the structure of operators 
in the cohomologies of the BRST charge shifted by the appropriate analytic solutions
in open string field theory ~{\sftmartin}. The obvious advantage of this approach is 
the background independence,
which in theory may allow us to penetrate 
beyond the realm of standard string perturbation theory.
Although at this time our understanding of how the analytic solutions work
to describe higher spin interaction is still very preliminary and limited,
in the end it seems plausible that the language of 
shifted BRST cohomologies in open string field theory
may be the most natural and efficient
 to understand the nonlocalities in higher spin interactions.

\centerline{\bf 6. Acknowledgements}

It is a pleasure to thank Lars Brink and other organizers and participants
of the International Workshop on Higher  Spin Gauge Theories at Nanyang Technological 
University, Singapore in November, 2015 for hospitality and inspiring discussions.
This work is supported by the National Natural Science Foundation of China under grant 11575119.

\listrefs

\end